\documentclass[useAMS,usenatbib]{mn2e}

\input epsf
\usepackage{graphicx}

\title[3D hydrodynamical simulations of the large scale structure of W50-SS433]{3D hydrodynamical simulations of the large scale structure of W50-SS433}
\author[Zavala et al.]{Jes\'us Zavala$^{1}$\thanks{E-mail:
jzavala@nucleares.unam.mx. Present address: Shanghai Astronomical Observatory, Nandan Road 80, Shanghai 200030, China; jzavala@shao.ac.cn}, Pablo F. Vel\'azquez$^{1}$\thanks{Email: pablo@nucleares.unam.mx}, Adriano
H. Cerqueira$^{2}$\thanks{Email: hoth@uesc.br}, Gloria
M. Dubner$^{3}$\thanks{Email: gdubner@iafe.uba.ar. Member of the Carrera del
  Investigador Cient\'\i fico del CONICET (Argentina)} \\
$^{1}$Instituto de Ciencias
Nucleares, Universidad Nacional Aut\'onoma de M\'exico, Ciudad
  Universitaria,\\
 Apartado Postal 70-543, CP 04510, M\'exico
D.F., M\'exico.\\
$^{2}$LATO-DCET-UESC, Rodovia Il\'eus-Itabuna, km 16 - 45662-000, Ilh\'eus,
  BA, Brazil.\\
$^{3}$Instituto de Astronom\'\i a y F\'\i sica del Espacio, CC 67  Suc 28,
(1428) Buenos Aires, Argentina.}

\begin{document}

\date{Accepted . Received ; in original form }

\pagerange{\pageref{firstpage}--\pageref{lastpage}} \pubyear{2002}

\maketitle

\label{firstpage}

\begin{abstract}

We present 3D hydrodynamical simulations of a precessing
jet propagating inside a supernova remnant (SNR) shell, particularly applied
to the W50-SS433 system in a search for the origin of its peculiar elongated
morphology.
Several runs were carried out with different values for the mass loss rate of the jet, the
initial radius of the SNR, and the opening angle of the precession cone. 
We found that our models successfully reproduce the scale and morphology
of W50 when the opening angle of the jets is set to 10$\degr$ or if this angle linearly
varies with time. For these models, more realistic runs were made
considering that the remnant is expanding into an interstellar medium (ISM)
with an exponential density profile (as HI observations suggest). 
Taking into account all these ingredients, the large scale morphology of the
W50-SS 433 system, including the asymmetry between the lobes (formed by the 
jet-SNR interaction), is well reproduced. 

\end{abstract}


\section{Introduction}
\label{sec:intro}

The standard supernova remnant (SNR) evolution theory, 
 proposed by \citet{wolt70,wolt72}, predicts a 
spherical morphology for remnants expanding into a uniform ISM.
Observations show, however, that a spherical morphology is not
the prevalent appearance of SNRs. Non-spherical 
morphologies can be produced 
by external or internal factors. Among the external factors are the 
existence of dense clouds or strong density gradients in the ISM 
where the SNR is 
expanding. Internally, in remnants resulting from a Type II SN explosion,
a compact object (a neutron star or a pulsar for example) can remain inside 
the SNR, injecting enormous amounts of energy through its magnetic
field and particle outflow, thus modifying the evolution of the host SNR.

Based on astronomical observations (mainly in radio continuum and X-ray), SNRs 
can be classified in three main morphological groups: (a) shell-type,
when the remnant exhibits a shell morphology in both radio and X-ray images;
(b) plerions or Crab-like type, when the morphology of the remnant resembles 
the appearance of a centre-filled nebula, similar to the Crab's SNR
(in this case the X-ray
and radio emission mainly come from the central compact object and the 
SNR shell is not observed); and (c) composite, when the SNR has a shell-type 
morphology plus a central component. The thermal-composite or mixed SNRs
belong to this group; they are characterised by a radio shell with
thermal X-ray emission in the interior. 

There are remnants with more complex morphologies that can be included in
group (c); such is the case of W50 \citep{Geld80,downes,elston87}.  The
morphology of this remnant can be described as an almost spherical shell
distorted by two symmetrical lobes along the East-West direction (see for
example fig. 1a and 1b of Dubner et al. 1998).  The source of relativistic
jets SS433 \citep{abell79,fabian79,clark78,hj82,margon84,hj85} is located at
the centre of W50.  These jets have a speed of 0.26c and a precession period
of 164 days and form a cone around the source with a half-angle of
$\sim$20$^{\circ}$. Analysis of the nature of the system reveals that the
source of SS433 is a binary system, with an orbital period of 13 days; the
jets' axis has an additional nodding motion with a period of 6 days
(Newsom \& Collins 1981, see also Collins \& Scher 2002). The precession axis
is aligned with the east-west direction of the W50 lobes. Several
authors have given different explanations for this alignment. These
explanations can be
grouped in two possible scenarios: first, the SS433-W50 system is formed by
the interaction of the jets of SS433 with the ISM (homogeneous or previously
swept up by the stellar wind of SS433), without being affected by the SNR
\citep{konigl,begelman}; second, the eastern and western lobes of the system
were formed by the interaction of the jets with the SNR shell
\citep{zealey,downes,murata}.  Observationally, \citet{elston87} found
evidence supporting the latter idea. Later on, \citet{dub98} also found
evidence based on high resolution radio-continuum images obtained with the
Very Large Array (VLA).  This was recently confirmed by the X-ray XMM-Newton
observations of \citet{brink07}.

A recent work carried out by \citet{ph01} (and references therein)
examines SNR catalogs looking for other possible examples of SNRs with
a similar elongated appearance. They identify three possible
candidates: G309.2-00.6, G41.1-0.3 located in the 3C397 complex, and
G54.1+0.3.

Following the jet-SNR interaction scenario, \citet{vel00} carried out 2D 
axisymmetric numerical simulations to describe the W50-SS433 
system, obtaining a rough agreement between their model and observations. 
Their model, however, was constrained by several hypothesis and
assumptions.  

In this work, we continue the study of \citet{vel00}, but
remove several of their assumptions. We present 3D numerical simulations for
the W50-SS433 system with the objective of providing a more realistic 
approach than the 2D model, following carefully the dynamics of the system as 
the
precessing jets inject energy and momentum into the SNR cavity, 
collide with the SNR shock front, and change the spherical shape of the 
remnant. These simulations were carried out with the Yguaz\'u-a code 
\citep{raga00,raga02}.

The article is organised as follows. Section 2 gives the initial conditions
of the numerical simulations and a brief description of the Yguaz\'u-a code. 
The results are presented in Section 3. Finally, the discussion and
conclusions of our work are given in Section 4.

\section{Initial conditions and the Yguaz\'u-a code}

\begin{table}\centering

  \caption{Parameters for the numerical simulations. {\em lin} 
means $\alpha(t)=\omega_1 t$, see text for details.}
  \begin{tabular}{lccccc}
\hline\hline
    Model & M1 & M2 & M3 & M4 & M5 \\
\hline
 $\dot{M}_j(10^{-6}M_{\odot}yr^{-1})$& 5 & 100 & 5 & 5 & 5 \\
 $\alpha$ & 20 & 20 & 20 & 10 & 20({\em lin}) \\
 $R_{SNR}(pc)$  & 20 & 20 & 10 & 20 & 20 \\
\hline
  \end{tabular}\label{table1}
\end{table}

As input data, we will use known values of the initial SN energy explosion,
jet velocity, density, precession period and angle, etc., based on the only
well-analysed observed case: the interaction between the relativistic jets
from  SS433 with the W50 SNR.  Previous works have used 2D
axisymmetric models of conical jets to describe this problem
\citep{vel00,ko90}, because the 164 day precession period of the jet 
\citep{margon79,hj88,hj85,hj82} is considerably
shorter than its dynamical evolution time, defined as
$t_{dyn}=R_l/v_j\simeq 140$ yr. Nevertheless, several assumptions on the
geometry of the jet are needed in this type of model. 2D axisymmetric
simulations model the symmetry axis as a reflecting boundary. This condition
can produce undesirable effects such as an increase of the momentum flow along
the symmetry axis \citep[see for example][]{raga07}.  Furthermore, in 2D
axisymmetric simulations, it is not clear if the precession cone should be
modelled by a filled conical jet or a hollow conical jet to represent the
actual physical processes that take place in such a system. With a full 3D
simulation, the assumption of axisymmetry is no longer needed and problems
regarding the reflection condition disappear.

We have carried out numerical simulations of five models for precessing jets
evolving into a SNR cavity. The main parameters for these models (the
half-angle of precession $\alpha$, the initial SNR radius $R_{SNR}$ and the
jet's mass injection rate $\dot{M}_j$) appear in table \ref{table1}.

The numerical simulations were made employing the 3D version of the Yguaz\'u-a
code (see \citet{raga00} and \citet{raga02}). The main characteristic of the
code is that it has an adaptive grid and integrates the gas-dynamical
equations with a second order accuracy (in space and time) implementation of
the ``flux-vector splitting'' algorithm of \citet{vanleer}.  Together with the
gas-dynamical equations, a system of rate equations for atomic/ionic species
is also calculated, which are then used to compute a non-equilibrium cooling
function (a parameterised cooling rate being applied for the high temperature
regime). They are also integrated with the gas-dynamical equations. Rate
equations for H I-II, He I-III, C II-IV, and O I-IV are considered (for more
details about reaction and cooling rates see Raga et al. 2002).

In all simulations, a 5-level binary adaptive grid was used with a maximum
resolution of $4.8\times10^{17}$~cm. A Cartesian computational domain of
$512^3$ pixels with a size of 80~pc along each axis was employed. The centre
of the SNR and the source of the jets is located at the centre of the $xy-$
plane with $z=0$. Free outflow conditions were employed on each computational
domain boundary.

The initial conditions for all models were adopted from different
multiwavelength observational studies. For the jet, we adopt velocity
$v_j=7.5\times10^9 {\rm cm\ s^{-1}}$, temperature, $T_j=1\times10^7{\rm K}$,
radius, $2.4\times10^{18} {\rm cm}$ (equivalent to 5 pixels for the maximum
resolution) and length, $4.8\times10^{18} {\rm cm}$. For the ISM we adopt
density, $0.01 {\rm cm^{-3}}$ \citep{safi97}, temperature, $1\times10^4 {\rm
  K}$ and velocity equal to zero; finally, the SNR was initialised employing
the autosimilar Sedov solution \citep{sedov59} considering a remnant radius of
$6.2\times10^{19}$~cm or 20~pc (except for the model M3 that has half of this
radius) and a SN explosion energy of $5\times10^{50}$~ergs.

\subsection {The ISM density gradient}

It is known that the system W50-SS433 is embedded in an ISM with a strong
density gradient \citep{dub98}, unlike the constant density ISM we have
assumed so far.  Thus, we made other runs of models M4 and M5 including a
stratified ISM. In this case the Sedov solution is not valid, so
 we divide the simulation in two parts. In the first one, the SN
explosion is simulated considering that an initial energy of $5\times10^{50}\ 
\mathrm{ergs}$ (the same energy used in the previous models) is put into a
sphere with a radius of 1.5~pc and a mass of 3 $M_{\odot}$, uniformly
distributed within this radius. Actually, the details of the SN explosion are
not important if our interest is to study an evolved SNR, such as W50, because
when the remnant enters into the adiabatic or Sedov evolution phase, the SNR
has already lost the memory of the initial conditions. The only important
parameter is the initial explosion energy $E_0$. The SNR evolution is followed
until the remnant shell reaches a radius of $\simeq 21$~pc, at an integration
time of $t=6000$~yr. In the second part, the jets are ``turned on'' within the
obtained SNR shell using the same parameters as in the previous models.

For these new runs, a
computational domain of $512\times 512\times 1024$ pixels was used,
corresponding to a physical size of $80\times 80\times 160$pc. Both SNR
and source jets were located at the centre of the computational domain. Free
outflow conditions were applied on each grid boundary.

The density stratification is given by
$\rho(x,z)=\rho_0$ exp$(-z_p/h_z)$, where 
$z_p=x~\textrm{sin}\beta-z~\textrm{cos}\beta$, $h_z=20$pc is a 
characteristic scale length, and $\beta=20\degr$ is the inclination of the
precession cone axis with respect to the $z-$~axis. 

\section{Results}

\begin{figure}
\includegraphics[width=\columnwidth]{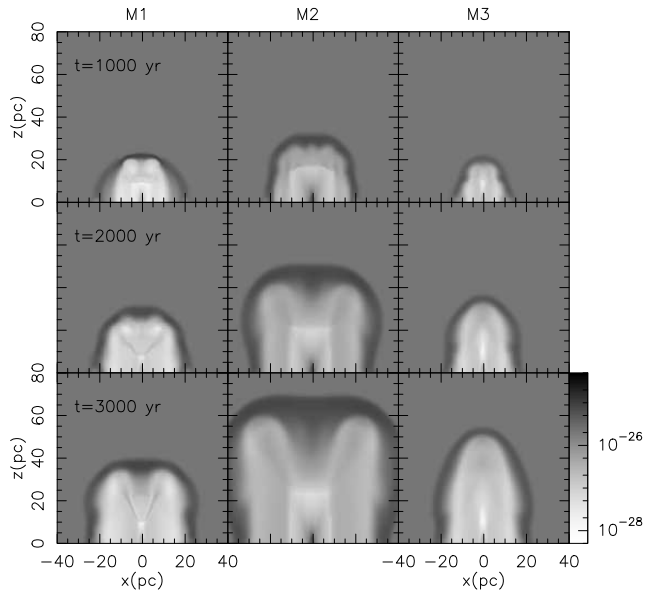}
  \caption{Comparison of the temporal evolution of density stratification maps for models M1, M2 and M3
    (left to right). The grey scale is logarithmic and the density is given in
    units of $gr\ cm^{-3}$.} 
 \label{3models}
\end{figure}

As a first analysis, we study the interaction of the precessing jets
with the shell of the SNR for $\alpha=20^{\circ}$. With the objective of 
reproducing the observed morphology of W50, we explore different 
values for the jet's mass injection rate (models M1 and M2) and the SNR 
radius (models M1 and M3). The 
range used for $\dot{M}_j$ lies in the range inferred from X-ray observations
\citep{safi97,kotani06}. The time delay between the SN
explosion and the ``turn on'' of the jets is not well-known; this
motivated us to make a run with a shorter SNR radius to simulate a
shorter time delay (model M3). 

\begin{figure}
\includegraphics[width=\columnwidth]{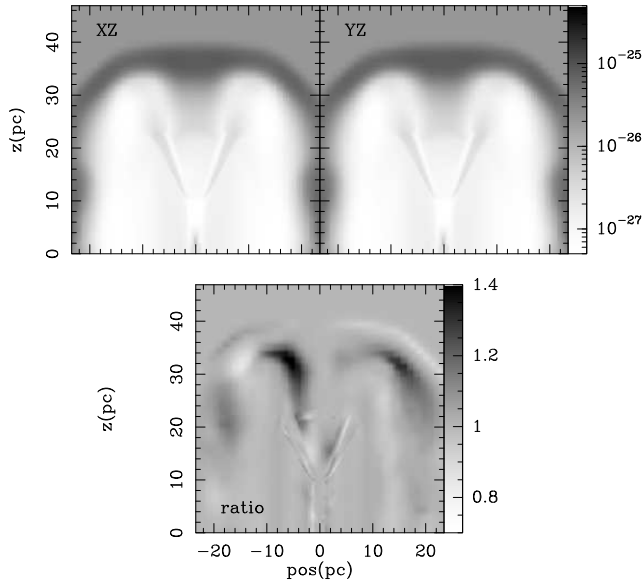}
  \caption{Asymmetry between xz and yz projections of the density distribution
    at $t=3000$ yr for model M1. Bottom panel displays the ratio between the
    $xz-$ and $yz-$ density distribution maps. If the system were perfectly
    symmetric, this ratio would be 1. Note that some regions show asymmetries
    with values $\sim$40\% showing the importance of full 3D hydrodynamical
    simulations to model the system.}
 \label{asm}
\end{figure}

Figure \ref{3models} shows the temporal evolution of the density distribution
for these models. The maps were obtained making a cut in the $xz-$ plane
passing through the centre of the jet's source. \citet{ko90} reported that
hollow conical jets, expanding into a uniform ISM, experience a
geometrical dilution of the momentum flux which prevents the jet's expansion.
A similar situation occurs in models M1-M3. Even though we are not using the
hollow conical jet approximation, we expect that the symmetric
approximation will be valid for distances sufficiently close to the source.  
The asymmetry in the density distribution due to the jet's precession can be
appreciated by zooming on the central region; this is shown in fig.
\ref{asm} for model M1. The left and right panels display the $xz-$ and $yz-$
density distributions, respectively.  The bottom panel is the point to point
ratio between these two projections, and it clearly shows the asymmetry of the
system, being larger for larger distances to the source. Even when the axial
asymmetries are not very large, the figure shows the advantage of a 3D
simulation over the usual assumption of axial symmetry.

Comparing models M1 and M2, it is clear that the increase of $\dot{M}_j$ does
not change the general structure of the cavity shell; the main difference is 
the larger cavity size for model M2. In the case of model M3, with
$R_{SNR}=10$~pc, the resulting morphology is very elongated along the jet 
axis, and the SNR shell quickly loses its identity.

In these models, the jet is moving into a stratified medium (the SNR
cavity) which could bend the initial jet's trajectory. However,
a deflection of the jet is not observed. We conclude that a precessing jet with
$\alpha=20^{\circ}$ moving into a remnant cavity cannot reproduce the
observed shape of W50.

Because  it was not possible to reproduce the morphology of $W50$ with
models M1-M3, we ran a new simulation, model M4, with the same parameters
as M1 but with $\alpha=10\degr$. 

The left panels of figure \ref{M4} display the temporal evolution of the
density stratification for model M4. By $t=1000$~yr, the SNR shell has been
broken by the precessing jet. The evolution snapshots show that the jet has
suffered a collimation process caused by the interaction of the jet's bow
shock and the SNR shell. As a result, secondary shocks are formed which
further reflect themselves on the $z-$~axis. For $t=3000$~yr, we can see that
there is a rupture of the primary lobe present at $t=2000$~yr. This pattern
resembles the helical structure shown in the eastern lobe of W50.
The result is also in agreement with the 2D axisymmetric modelling of
 \citet{vel00}.

\subsection {The semi-aperture angle of the jets}       

These results point to an apparent contradiction between the observed
semi-aperture angle $\alpha$ and our numerical models. How can they be made
compatible?. Previous X-ray observations
\citep{watson83,brink96,safi97} have shown that the angle subtended by the
``X-ray lobes'' is smaller than expected from the precession cone
of the innermost SS433 jets (40\degr). A possible solution  is to
consider that $\alpha$ is not constant during the jet evolution. We
will use a model where $\alpha$ starts with a value equal to zero and then
grows with time.  We assume a simple model where $\alpha$ depends linearly
on time: $\alpha=\alpha_m\omega t$ for $0\leq t\leq\tau_c$ and
$\alpha=\alpha_m$ for $t>\tau_c$, where $\omega=1/\tau_c$, $\tau_c$ is a free
parameter that was chosen to be 2000 yr in order to produce lobes with the
desired size, and $\alpha_m=20\degr$ is the estimated value of the opening
angle today \citep{abell79}.

\begin{figure}\centering
\includegraphics[width=7.5cm]{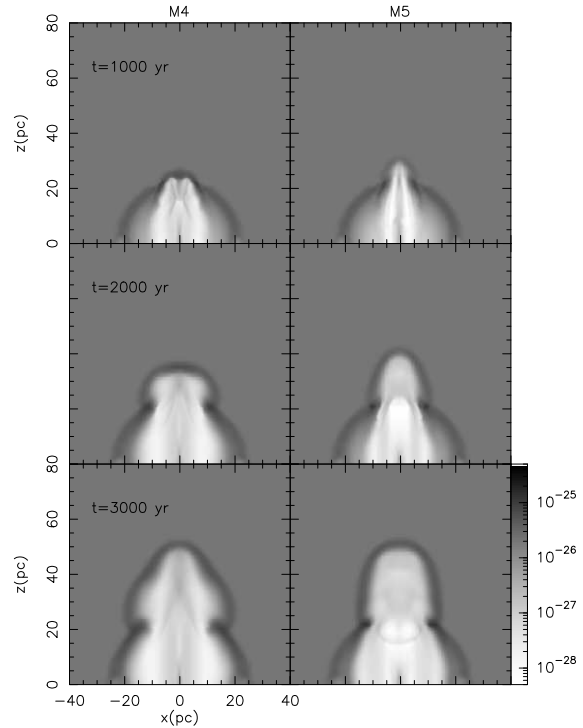}
\caption{Snapshots of the density distribution for models M4 (left
  panel) and M5 (right panel) at integration times $1000$~yr,
  $2000$~yr and $3000$~yr. Note that the spherical symmetry of the SNR
  is broken by the precessing jet and the formation of the lobe is
  clearly observed.}
  \label{M4}
\end{figure}

This hypothesis is included in model M5 and the results are shown
in the right panels of figure \ref{M4}. The snapshots are for the same
evolution times as the left panels.
The shock front of the jet breaks the SNR shell as early as $t=1000$~yr and for
$t=3000$~yr a well-formed lobe is observed. Unlike model M4, there
is no secondary breaking of the lobe.

\subsection {The ISM density gradient effect}

\begin{figure}
\includegraphics[width=8.5cm]{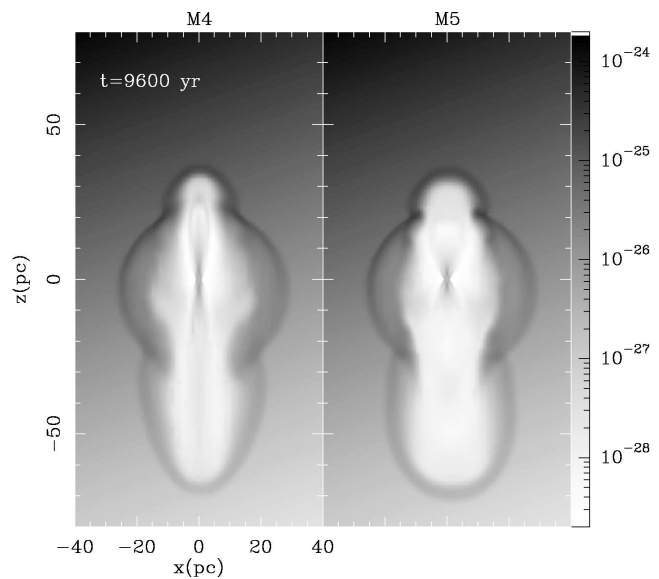}
  \caption{The same parameters for the jet and SNR as for models M4 (left
    panel) and M5 (right panel) but the ISM has a density gradient increasing
    along an axis tilted by an angle of $20\degr$ with respect to the vertical
    axis. The snapshot is for $t=3600$~yr after the jet is turned on (a total
    evolution time of 9600~yr).}
  \label{w50dens}
\end{figure}

The effects from the inclusion of the ISM density gradient to our
model, described in subsection 2.1, can be seen in figure
\ref{w50dens}. The left panel shows the density distribution on the
$xz$ plane for a run with the same parameters as for model M4, while
the right panel has the same parameters as model M5.  These maps were
obtained for an integration time of 3600~yr after the jet is turned on
(corresponding to a total evolution time of 9600~yr).  The ISM density
gradient produces an extra asymmetry in the system which is clearly
seen by the difference in the evolution of the lobes corresponding to
the upper (western) and lower (eastern) jets. In the region where the
ISM density is higher, the effect is to slow the SNR shell evolution
and also to slow the penetration of the SNR shell by the jet's cocoon.

\begin{figure}
\includegraphics[width=8.5cm]{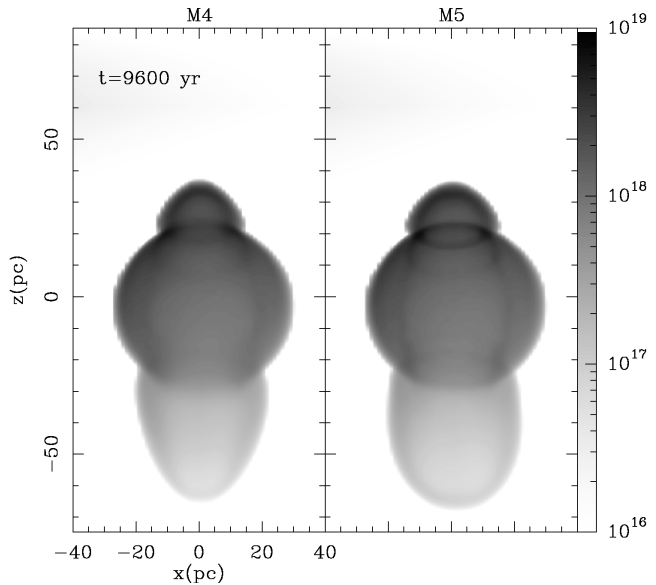}
  \caption{Electronic density integrated along the line of sight with a 
    $21\degr$ inclination angle with respect to the plane of the sky for the
    same models as figure \ref{w50dens}. The grey scale is logarithmic and is
    given in units of $\mathrm{cm}^{-2}$}
  \label{w50ne}
\end{figure}

In order to have a more direct comparison with the actual observed system, we
produced maps of the electronic density integrated along the line of sight for
the new runs of models M4 and M5. These maps are shown in the left and right
panels of figure \ref{w50ne}, respectively. An inclination angle of $21\degr$
with respect to the plane of the sky has been taken into account.  The angle
was chosen according to the observed value given by \citet{hj88}.  The
electronic density is a measure of the ionised zones in the system, it is
expected that it will be enhanced just behind the shock fronts. For this
reason, the electronic density is a good tracer of the structure observed in
radio-continuum images. The physical sizes of the resulting distorted SNR
shell are $56\mathrm{pc}$ and $103\mathrm{pc}$, along the $x-$ and $z-$
directions, respectively. These physical sizes correspond to angular sizes of
$1\degr 4\arcmin \times 1\degr 58\arcmin$ (if a distance of 3~kpc to SS433 is
assumed), which are very similar to the observed dimensions ($1\degr \times
2\degr$).  The main difference between the panels of figures \ref{w50dens}
and \ref{w50ne} is that the more evolved lobe (lower) is more collimated for
the case of model M4 than for M5.

Figure \ref{w50comp} illustrates the comparison between simulations
(left panel of fig. \ref{w50ne}) and radio observations.  The
similarity with the system W50-SS433 is appealing. The eastern (left)
lobe in the simulation is however wider than the observed one. A
possible cause for this difference is the presence of an HI cloud in
the southern part of the eastern lobe \citep[as suggested by][]{dub98},
which could have contributed to an extra collimation of the
lobe. Since this cloud is not included in our models, the result is a
wider lobe. It is important to point out that an ideal comparison
would be with synthetic radio maps generated directly from the
simulations.  However, our models do not include magnetic fields and
thus it is not possible to produce such maps.

\section{Summary and Conclusions}

\begin{figure*}
\includegraphics[width=16cm]{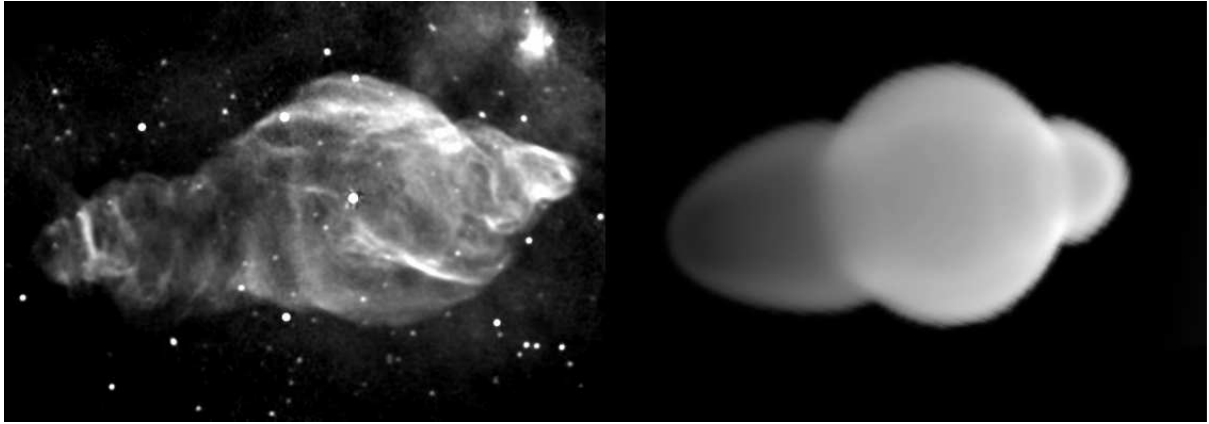}
  \caption{Comparison of the radio-continuum image with the simulated column
    electronic density map of model M4. The left panel shows the 1415 MHz
    image in grey scale and contours of the W50 SNR, obtained with the VLA by
    \citet{dub98}, in equatorial coordinates (North is up). The right panel
    shows the simulated map in a grey color scale. A distance of 3kpc to 
   SS433 was assumed.}
 \label{w50comp}
\end{figure*}

We have analysed the interaction between a precessing jet and a SNR
shell using a 3D hydrodynamical simulation with the goal of 
studying the large scale structure of the SS433-W50 system. The motivation for
doing so is to put aside one of the usual assumptions made in the past
regarding this kind of system, namely, using a cone (hollow
or filled) as an approximation to model the jets from the source
SS433. The precession period of the jets is lower than the dynamical
time of the system, therefore, it would be expected that the
precessing jets can be modelled by a single axially symmetric conical
jet. Nevertheless, analysing our simulations, we found that there is a
significant axial asymmetry, especially for material located far away
from the central source ($\sim20-40\%$ of the ratio between the $xz$
and $yz$ density projections, see fig. \ref{asm}).  These results
confirm the advantages of using 3D over 2D axisymmetric simulations.

We found that the simple simulated model with a constant value 
for the aperture angle, as reported observationally for the jets
in SS433: $\alpha=20\degr$, fails to reproduce the observed morphology of this
system. This result holds in spite of exploring a large range of values for 
the mass loss rate
and the initial radius of the SNR (models M1-M3, see fig. \ref{3models}). This
radius is connected with the time delay between the SN explosion and 
the ``turn on'' of the jets. If this delay is too short, the jet's cocoon 
sweeps up the SNR shell erasing all traces of the initial spherical morphology
(model M3, right panel fig. \ref{3models}).

If the value of $\alpha$ is arbitrarily reduced by half (model M4, see
left panel of fig. \ref{M4}), then the shock front of the jets is
collimated enough to break the SNR shell, forming two lobes very
similar to those observed. This result is related to the observational
fact that at a distance of $\sim6\times10^{19}$cm from the source
center, the angular extent of the X-ray jet is $\sim18\degr$
{\citep[see for instance,][]{brink07}}, far below the expected
$40\degr$ of the original model. This suggests a re-collimation
mechanism that is still unknown. It is no surprise then that the
large scale morphology of the system is reproduced well by our model
when the opening angle is artificially reduced by half.

Such results led us to explore a different model where the value of $\alpha$
changes with time (model M5, see right panel of fig. \ref{M4}), starting
from zero and growing linearly until it reaches the constant value of
$\alpha=20\degr$. This model also successfully  reproduces the expected
morphology.  It is interesting to note that the corkscrew patterns observed in
radio maps, at the tips of both Eastern and Western lobes of W50 (possibly
produced by precession of the arcsecond-scale jet from SS 433) are limited
spatially by a cone which seems to be smaller than the derived $\sim
20^{\circ}$ half-opening angle of the sub-arcsecond-scale jet. This is clearly
illustrated in figure 1 of \citet{moldowan05}, where a $\sim 20^{\circ}$
half-opening cone is superimposed on the radio map of W50. The authors also
suggest that the sub-arcsecond jet is playing a crucial role in determining
the morphology of W50 nebula, since their Chandra observations of the inner
region of the west lobe of W50 show a good correlation between the precession
of the jet and the X-ray emission.  Based on this evidence, we
presented here a numerical exercise, in order to investigate the morphological
effects of a time varying opening angle jet on the whole structure of the
nebula.

Finally, in order to give a more realistic model, two more simulations were
made by using the parameters of models M4 and M5 but taking into account
the observed density gradient of the ISM surrounding W50 \citep{dub98}. In
these runs the Sedov solution for the SNR was not used, instead, the
simulations were divided in two stages: in the first one the SNR evolves
alone; the second part follows with the turn on of the jets. The effect of the
ISM density gradient is to produce different evolutionary patterns for the two
lobes, making one of them more developed (see fig. \ref{w50dens}).
Given the fact that in our description the magnetic field is not included, it
is not possible to generate synthetic radio continuum maps; an alternative was
to present maps of the electronic density integrated along the line of sight,
which is a good tracer of the shock fronts. An inclination of the system with 
respect to the plane of the sky of $\sim21\degr$ was considered to obtain this
map (see fig. \ref{w50ne}). 

In regard to the scale of the system, our models give physical sizes
of $56\mathrm{pc}\times 103\mathrm{pc}$, which are similar to the
observed dimensions, if a distance of 3~kpc is considered for W50
\citep[based on the HI study of][]{dub98}. There are other distance
estimations for this remnant in the literature.  Recently,
\citet{lockman07} gives a distance between 5.5 and 6.5 kpc. This
however does not change our main results, it would only be necessary
to re-scale the simulated sizes and times.

In summary, the presented 3D models of the precessing jets of SS433
interacting with the shell of W50 reproduce the global characteristics
(morphology and size) of this astrophysical object (see fig. \ref{w50comp}),
giving a closer qualitative and quantitative agreement with observations. It
is however important to say that the model is still far from being complete;
it is of key importance to propose and test a model that can account for the
early re-collimation of the jets. The magnetic field could be included in
future works as a possible solution to this problem (this was analytically
treated by Kochanek 1991, but for conical jets moving into an uniform ISM).
Other details of the structure of the system, such as the reported bend in the
eastern jet's propagation direction (see for example \citet{brink07}) can be
studied by simulating a more complex ISM. Also, future works could focus on
the late evolution of the jets, when they are far from the central regime.

\section*{Acknowledgments}

The authors thank the referee for her/his comments, which helped us to
improve the previous version of this manuscript.  The authors
acknowledge support from CONACyT grants 46628-F, and DGAPA-UNAM grant
IN108207. JZ acknowledges CONACyT and DGEP (UNAM) scholarships, and support by the CAS Research Fellowship for 
International Young Researchers. The
work of PFV was supported by the ``Macroproyecto de Tecnolog\'\i as
para la Universidad de la Informaci\'on y la Computaci\'on''
(Secretar\'\i a de Desarrollo Institucional de la UNAM, Programa
Transdisciplinario en Investigaci\'on y Desarrollo para Facultades y
Escuelas, Unidad de Apoyo a la Investigaci\'on en Facultades y
Escuelas). GMD is thankful for financial support from grants
PIP-CONICET(Argentina) 6433 and PICT-ANPCYT(Argentina) 14018. We thank
Stan Kurtz (CRyA, Morelia, UNAM) for reading this manuscript. We also
would like to thank the computational team of ICN: Enrique Palacios
and Antonio Ram\'\i rez, for maintaining and supporting our Linux
servers; and Mart\'\i n Cruz, for the assistance provided.

\bsp
\label{lastpage}
\end{document}